\documentclass[prl,aps,showpacs,twocolumn]{revtex4}
\usepackage{graphicx}
\usepackage{dcolumn}

\begin{document}


\title{

Predominance of Prolate Nuclear Deformations Emerging from
Many-Body Interactions}

\author{Mihai Horoi}
\affiliation{Department of Physics, Central Michigan University,
Mount Pleasant, Michigan 48859, USA}
\author{ Vladimir Zelevinsky}
\affiliation{
National Superconducting Cyclotron Laboratory,
East Lansing, Michigan 48824, USA, and
Department of Physics and Astronomy, Michigan State University, East
Lansing, Michigan 48824, USA}

\begin{abstract}

A new approach to the old problem of the predominance of prolate deformations
among well deformed
nuclei is proposed within the shell model framework. The parameter space is
explored using the ensemble of random rotationally-invariant
interactions. Subsets with rotational energy ratio $E(4^{+})/E(2^{+})$
and the rigid-rotor relation between the quadrupole moment $Q(2^{+})$
and the transition probability $B(E2;2^{+}\rightarrow 0^{+}))$ are found 
exhibiting prolate predominance.
We identify matrix elements of the effective forces responsible
for the predominance of prolate deformation.

\end{abstract}

\pacs{21.60.Cs, 21.10.Re, 21.60.Ka}

\date{\today}
\maketitle



The majority of deformed nuclei have axially symmetric prolate deformation in
their ground state (g.s.). The presence of rotational bands with typical energy
intervals is the first signature of stable deformation. The states in a band
are connected by strong quadrupole transitions obeying simple rigid rotor
intensity rules. The same rules \cite{BM89} determine the expectation values of
multipole operators. This is essentially a projection of nearly constant within the band
intrinsic (body-fixed) quantities onto the laboratory (space-fixed) coordinate
frame. The overwhelmingly positive sign of the intrinsic quadrupole moment
manifests the dominance of prolate deformation. The Nilsson diagram shows a
pattern of split spherical single-particle orbitals that seems to be more or
less symmetric with respect to the sign of deformation; the difference in
the liquid drop energy for the two signs of deformation is pretty small \cite{mottprivate}.
While the results of
the numerous mean field calculations for specific nuclei in general correctly
predict the presence and the sign of deformation, the underlying physical reason for
the predominance of prolate shapes is unclear.

It was argued by Lemmer \cite{lemmer60} that the kinetic energy should contain
an additional contribution similar to the term $-D\vec{ \ell}^{2}$ introduced
by Nilsson \cite{nilsson55} for interpolation between the harmonic oscillator
and potential box. Being split to time-conjugate pairs $(m,-m)$ by
deformation, the large-$\ell$ spherical orbitals determine the single-particle (s.p.)
occupancies in the way that makes the prolate case energetically favorable.
Related arguments were given by Castel and Goeke \cite{castel76}, who
showed that the collective
inertial parameter is larger for the prolate deformation, and later by Castel, Rowe
and Zamick \cite{castel90} with the help of self-consistency conditions \cite{BM89}.
There are also ideas
based on the semiclassical analysis of the s.p. level density
\cite{frisk90}, simplest periodic orbits \cite{deleplanque04} and their
bifurcations in a deformed cavity \cite{arita98}. The predominance of prolate
deformation analogously emerges in s.p. motion in metallic clusters
\cite{hamamoto91} and for many biological objects, from molecular level to
pollen grains \cite{pollen03}. The latest analysis by Hamamoto and Mottelson
\cite{hamamoto09} starting with the statement that ``the nature of the
parameters responsible for the prolate dominance has not yet been adequately
understood"  proceeds along similar lines. Detailed calculations and comparison
between the harmonic oscillator potential and that of a spheroidal cavity show
a different character of the mixing of spherical orbitals for the prolate and
oblate cases. Being a surface effect, the difference should decrease in large systems.
The authors of Ref. \cite{hamamoto09} mention additional factors
which were not sufficiently accounted for including the role of the spin-orbit
potential, pairing effects \cite{tajima02}, and the presence of two kinds of
particles.


\begin{figure*}
\centering
\includegraphics[width=5.0in]{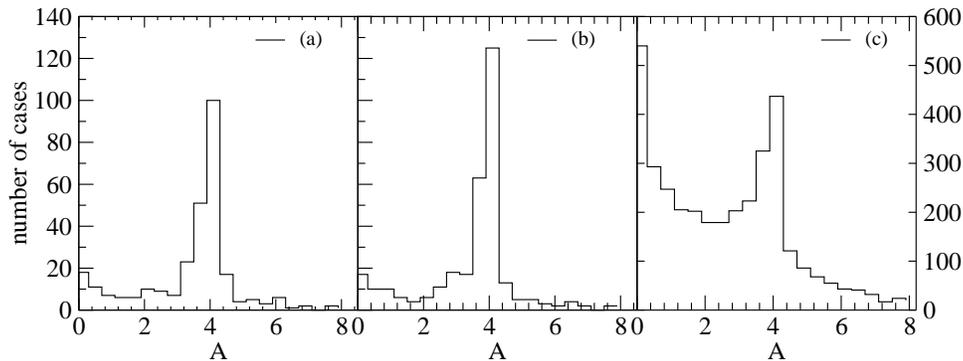}
\caption{
The distribution of the Alaga ratio, Eq. (\ref{1}), for all cases
selected by the energy ratio $E(4)/E(2)$; the interaction strength is
$\lambda=0.5$, panel {\sl a}, and $\lambda=4$, panel {\sl b}; the same for
all cases with the sequence $(0^{+},2^{+})$, panel {\sl c}.
}
\label{3panes}
\vspace*{-6mm}
\end{figure*}

Below we suggest a new approach to this problem that was previously attacked
from the viewpoint of the deformed mean field and corresponding
quasiparticle motion. As the mean field
itself is generated by the nucleon-nucleon interactions, it makes sense to take
a step back and consider the many-body problem rather than the resulting s.p.
motion. This would allow us to avoid questions of self-consistency and take
into account not only pairing and spin-orbit forces but the full
inter-particle interaction. The natural framework is
provided by the shell model (SM) with effective interactions; modern versions
of the SM give good agreement with the data.  The exact
diagonalization of the Hamiltonian matrix, with all conservation laws strictly
respected, provides the eigenstates in the space-fixed frame; the spectrum and
observables can be analyzed similarly to the experimental data. The problem
requires exploration of the parameter space of the SM Hamiltonians in a
sufficiently large orbital space.

To explore the parameter space we will work with random interactions. All
two-body matrix elements allowed by angular momentum, parity and isospin
conservation are taken as random uncorrelated quantities. Among many
realizations of such an ensemble we can identify those with definite rotational
characteristics and study the abundance of such cases and their dynamic origin.
Random interactions in many-body systems are usually analyzed for studies of
quantum chaos \cite{brody81,big} in relation to the random matrix theory. In
Ref. \cite{johnson98} the ensemble was constructed with rotationally invariant
interactions, and only the magnitudes of allowed matrix elements were taken
randomly. The unexpected output was the discovery of the predominance, for an
even number of particles, of the zero ground state spin, in spite of relatively
low multiplicity of $J=0$ states in Hilbert space. This result, valid not
only in a single-$j$ case but in the context of realistic SM space
\cite{HBZ01} as well, was fully
understood in terms of geometry of the parameter space only for the exceptional
case of the single $j=7/2$ shell \cite{chau02} and for the similar results in
the interacting boson model \cite{bijker02}. In more realistic fermionic models
the full theory is still absent; the most promising are the ideas of geometric chaoticity of
random angular momentum coupling \cite{mulhall00,papenbrock05}.
The details were discussed in the review articles
\cite{ZV04,zhao04,weiden07}.

The rotationally invariant random interaction models provide a wealth of information.
Any given set of random interaction parameters gives rise to a specific version of the
SM. It was noticed, for example, that
some sequences of the lowest excited states, like $0^{+}-2^{+}-4^{+}- ...$,
appear with the enhanced probability. The energy ratios in such yrast
sequences are spread around the values close to limiting cases of the
collective models, such as vibrational and rotational bands \cite{johnson07}.
In Ref. \cite{ZV04}, the
realizations with the lowest states $0^{+}-2^{+}$ were analyzed in terms of the
Alaga ratio,
\begin{equation}
A=\,\frac{Q(2^{+})^{2}}{B(E2;2^{+}\rightarrow 0^{+})},             \label{1}
\end{equation}
of the squared expectation value of the quadrupole moment of the $2^{+}$ state
to the reduced transition probability from this state to the g.s.
The distribution of the Alaga ratio reveals two
pronounced peaks, close to zero and at the value corresponding to the rigid
rotor \cite{BM89}. This means that the stable mean fields generated by random interactions
with high probability correspond either to near-spherical or to well
deformed shape. The situation is similar in the case of the interacting
boson model \cite{bijker02}.

We define the space-fixed quadrupole moment of the axially symmetric rotor with
spin $J$ \cite{BM89} as
\begin{equation}
Q(J)=Q_{0}\,\frac{3K^{2}-J(J+1)}{(J+1)(2J+3)}\,\Rightarrow \,-\,Q_{0}\,
\frac{J}{2J+3},                                                      \label{2}
\end{equation}
where $Q_{0}$ is the intrinsic (body-fixed) quadrupole moment, and we assume
$K=0$ for the yrast band of an even-even nucleus. Prolate intrinsic shapes,
$Q_{0}>0$, correspond to squeezed shapes around the axis of collective
rotation, $Q<0$. With the standard \cite{BM89} definition of the B(E2) probability for the yrast
band of a well deformed rotor, the Alaga ratio
(\ref{1}) is equal to 4.10.

\begin{table}
\begin{tabular}{c|c|c|c|c|c|c}
\hline
& $\lambda$ & $N(0,2)$& $\frac{N(Q<0)}{N(0,2)}$ & $N(E4/E2)$ &
$N_{{\rm rot}}$ & $\frac{N_{{\rm prolate}}}{N_{{\rm rot}}}$\\
\hline
$(a)$ & 0.05 & 1398 & 0.62 &  50 & 3 & 1.00 \\
& 0.5 & 3320 & 0.54 & 322 & 100 & 0.74 \\
& 1.0 & 3846 & 0.52 & 354 & 100 & 0.70 \\
& 1.5 & 4056 & 0.52 & 378 & 119 & 0.72 \\
& 2.0 & 4129 & 0.52 & 371 & 122 & 0.74 \\
& 3.0 & 4196 & 0.52 & 366 & 126 & 0.70 \\
& 4.0 & 4233 & 0.52 & 367 & 125 & 0.71 \\
& 10.0& 4295 & 0.53 & 368 & 112 & 0.74 \\
\hline
$(b)$ & 1.0 & 3156 & 0.55 & 289 & 39 & 0.77 \\
& 2.0 & 3153 & 0.53 & 264 & 34 & 0.79 \\
& 3.0 & 3140 & 0.52 & 266 & 34 & 0.82 \\
& 4.0 & 3156 & 0.53 & 240 & 35 & 0.89 \\
\hline
$(c)$ & 1.0 & 4569 & 0.54 & 322 & 120 & 0.73 \\
& 2.0 & 4530 & 0.52 & 339 & 116 & 0.75 \\
& 3.0 & 4490 & 0.52 & 349 & 119 & 0.76 \\
& 4.0 & 4461 & 0.52 & 371 & 124 & 0.81 \\
\hline
$(d)$ & 1.0 & 2170 & 0.43 & 176 & 55 & 0.07 \\
& 2.0 & 2185 & 0.37 & 193 & 70 & 0.06\\
& 3.0 & 2212 & 0.35 & 188 & 73 & 0.05 \\
\hline
\end{tabular}
\caption{The probabilities of occurrence of collective prolate configurations
(see text for details).}
\vspace*{-8mm}
\end{table}

In the single-$j$ model \cite{ZV04}, where the splitting of a $j$-level
proceeds fan-like with approximate symmetry of oblate and prolate sides, the
sign of $Q$ will be random. Therefore we expect the crucial role of mixing of
spherical orbitals. It was often suggested \cite{rmpsm} that the spherical SM
can realistically describe deformation if one uses at least two s.p. orbits
with $\Delta j = 2$. This approach successfully produces collective bands,
such as in $^{48}$Cr, and even in the double magic $^{56}$Ni \cite{ni56}. The
collectivity was tested not only by the $J(J+1)$ behavior, but also by large
and consistent in band quadrupole moments and $B(E2)$ strengths. As a generic
example, we consider a model with four neutrons and four protons in the space
of two spherical orbitals, $0f_{7/2}$ and $1p_{3/2}$, of the same parity.
This corresponds to the oversimplified SM description of $^{48}$Cr, the nucleus with
well known collective properties. Without
interactions, we would have half-filled $f_{7/2}$ orbitals with zero quadrupole moment.
In this space the most general rotationally
invariant two-body interaction is described by 30 matrix elements, Table II
below. We choose the random ensemble defined by the uniform distribution of
uncorrelated matrix elements between $-V$ and $+V$, and the only parameter is
the ratio $\lambda=V/\epsilon$, where $\epsilon$ is the (non-random) spacing
between the two orbitals. With the statistics of 10000 realizations for each
value of $\lambda$, the results are presented in Table I$a$. 

The case of very small $\lambda=0.05$ can be viewed as the single-$j$ result. The column
$N(0,2)$ shows the number of realizations with the sequence $0^{+},2^{+}$ of
the g.s. and the first excited state. The probability of such sequences is
strongly enhanced. Before applying additional requirements we can find the
sign of the expectation value (\ref{2}) of the quadrupole moment $Q(2^{+})$. As
expected, the fraction $N(Q<0)/N(0,2)$ of ``prolate" cases, $Q<0$, is stable
near 50\% (the third column). Among these states there are cases with
rotational properties. The constraint that the ratio $E(4^{+})/E(2^{+})$ be
between 3 and 3.6 (the rigid rotor would give 3.33) leaves the number of states
indicated in the column $N(E4/E2)$. In the column $N_{{\rm rot}}$ we count the
states where, in addition to the energy criterion, the Alaga ratio (\ref{1}) is
between 3.90 and 4.30, an arbitrary but restrictive choice. The last
column gives the fraction of states with negative sign of $Q$ among all
``rotational" states. The predominance of prolate configurations is practically
independent on the relative strength of mixing between the orbitals if this
strength is sufficient for the onset of deformation. The full histogram for the
distribution of the Alaga ratio is shown in Fig. 1 for two values of the
interaction strength; this figure shows also the statistics without any
rotational selection cuts. The prolate deformed peak is quite narrow. With the
only energy criterion $(0^{+},2^{+})$ applied, we see also a large peak for
spherical configurations with $A$ close to zero, as well as intermediate
background situations. In this large set, the prolate cases appear only in
slightly more than 50\% cases. We can also note that at very weak mixing,
$\lambda=0.05$, the numbers $N(0,2)$ and especially $N(E4/E2)$ fall sharply,
and the number of rotational configurations satisfying our criteria is very
small. This emphasizes the importance of orbital mixing.

\begin{figure}
\centering
\includegraphics[width=2.5in,angle=-90]{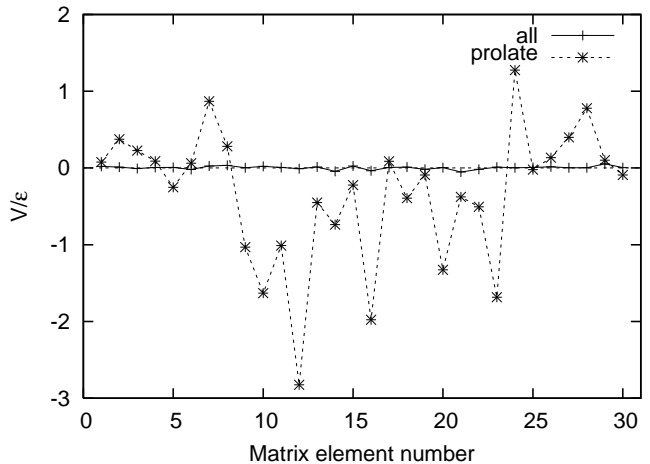}
\caption{
The average matrix elements of the two-body interaction.
The matrix element number is as in Table II.
}
\label{me}
\vspace*{-6mm}
\end{figure}

Going from the isospin-symmetric case with $N=Z$ to asymmetric one, we consider
next the system with six neutrons and four protons assuming the same set of two
s.p. levels. The results are shown by Table I$b$. The fraction of states with
$Q<0$ among all sequences $(0^{+},2^{+})$ is still close to 1/2. The total
number of rotational cases is now significantly lower. At the same time, the
fraction of prolate configurations is even higher than in the symmetric case.
The results are essentially the same in the $N=Z$ case for the inverted level
scheme when the spherical $p_{3/2}$ level is placed below the $f_{7/2}$ level,
see Table I$c$. Again for the very weak mixing, $\lambda=0.01$, we do not find
any rotational states; in this case all eight nucleons occupy the lowest
$p_{3/2}$ level creating the closed shell. With strong level mixing, 45\% of
all states belong to the $(0^{+},2^{+})$ sequence. Clearly, the occurrence of
prolate deformation is even higher than for the normal $fp$ level sequence.
This confirms the presence of the effect we mentioned in the introduction,
namely the influence of the splitting of the level with higher $\ell$ when
low-$|m|$ components steeply go down with the prolate deformation. The last
case we consider here contains four protons and four neutrons in the space of
two levels of opposite parity, $f_{5/2}$ and $g_{9/2}$, Table I$d$. Here the
fraction of the sequences $0^{+},2^{+}$ is lower, and practically all
rotational states have oblate deformation. It is clear that, due to parity
conservation, direct mixing of split levels by the deformed mean field is
impossible and only pairwise transfers of particles can play a role.

{\small
\begin{table}
\begin{tabular}{c|c|c|c}
& $\langle j_{1}j_{2}|V|j_{3}j_{4}\rangle(JT)$ & {\rm Full average} & {\rm
Prolate average} \\
 \hline
1 & $\langle ff|V|ff\rangle(10)$ & 0.021 & 0.078 \\
2 & $\langle ff|V|ff\rangle(30)$ & 0.012 & 0.374 \\
3 & $\langle ff|V|ff\rangle(50)$ & -0.007 & 0.227 \\
4 & $\langle ff|V|ff\rangle(70)$ & 0.007  & 0.089 \\
5 & $\langle ff|V|ff\rangle(01)$ & 0.008  & -0.252 \\
6 & $\langle ff|V|ff\rangle(21)$ & -0.020 & 0.062 \\
7 & $\langle ff|V|ff\rangle(41)$ & 0.026  & 0.869 \\
8 & $\langle ff|V|ff\rangle(61)$ & 0.034  & 0.282 \\
9 &  $\langle ff|V|pf\rangle(30)$ & 0.004  & -1.033 \\
10 & $\langle ff|V|pf\rangle(50)$ & 0.022  & -1.630 \\
11 & $\langle ff|V|pf\rangle(21)$ & 0.006  & -1.010 \\
12 & $\langle ff|V|pf\rangle(41)$ & -0.010 & -2.826 \\
13 & $\langle ff|V|pp\rangle(10)$ & 0.014  & -0.451 \\
14 & $\langle ff|V|pp\rangle(30)$ & -0.043 & -0.739 \\
15 & $\langle ff|V|pp\rangle(01)$ & 0.025  & -0.223 \\
16 & $\langle ff|V|pp\rangle(21)$ & -0.036 & -1.977 \\
17 & $\langle pf|V|pf\rangle(20)$ & 0.007  & 0.088 \\
18 & $\langle pf|V|pf\rangle(30)$ & 0.010  & -0.393 \\
19 & $\langle pf|V|pf\rangle(40)$ & -0.018 & -0.092 \\
20 & $\langle pf|V|pf\rangle(50)$ & 0.004  & -1.328 \\
21 & $\langle pf|V|pf\rangle(21)$ & -0.052 & -0.376 \\
22 & $\langle pf|V|pf\rangle(31)$ & -0.019 & -0.507 \\
23 & $\langle pf|V|pf\rangle(41)$ & 0.011  & -1.685 \\
24 & $\langle pf|V|pf\rangle(51)$ & -0.003 & 1.276  \\
25 & $\langle pf|V|pp\rangle(30)$ & 0.007  &-0.023 \\
26 & $\langle pf|V|pp\rangle(21)$ & 0.014  & 0.133 \\
27 & $\langle pp|V|pp\rangle(10)$ & 0.003  & 0.400 \\
28 & $\langle pp|V|pp\rangle(30)$ & 0.003  & 0.779 \\
29 & $\langle pp|V|pp\rangle(01)$ & 0.054  & 0.102 \\
30 & $\langle pp|V|pp\rangle(21)$ & 0.005  & -0.092 \\
\hline
\end{tabular}
\caption{List of average matrix elements for the case of Table I$a$ and  $\lambda=4.0$.}
\vspace*{-6mm}
\end{table}
}

The main new dimension brought in by our approach is in the visualization of the
entire parameter space. The mechanism of the prolate predominance is revealed as we identify the
realizations of the ensemble responsible for the effect.
Fig. 2 shows all 30 reduced matrix elements $\langle
j_{1}j_{2}||V||j_{3}j_{4}\rangle(JT)$ allowed by angular momentum and isospin for
the strong coupling limit, $\lambda=4.0$, in case $(a)$ of Table II.
The average value of each matrix element over the whole ensemble is zero, with
small fluctuations (solid line). The dashed line shows the average over prolate
rotational samples. Table II displays the ordering of the matrix elements and
their average numerical values. Although all matrix elements are enhanced
compared to the level of fluctuations, one can conclude that certain matrix
elements are crucial for the transition from spherical shape to predominantly
prolate axial deformation. We see the exceptional role of amplitudes {\sl
9}-{\sl 12} describing the transfer of a single nucleon $f_{7/2}\leftrightarrow
p_{3/2}$ in the interaction with another $f_{7/2}$ nucleon, regardless of its
isospin. Such a process is forbidden for the orbitals of opposite parity. This
agrees with the idea discussed in Ref. \cite{hamamoto09}. The monopole pairing
given by amplitudes {\sl 5},{\sl 15} and {\sl 29} is not effective in this
transition. Contrary to that, we see a large negative
amplitude of quadrupole pair transfer {\sl 16}. Large quadrupole-quadrupole
forces in the particle-hole channel correspond to matrix elements {\sl 20}-{\sl
24} in the particle-particle channel. They induce the collectivization process
and formation of the proper symmetry of the mean field after mixing $p$- and
$f$-orbitals.

In summary, for the first time we performed the exploration of the parameter
space that serves as an arena for effective nucleon-nucleon interactions
building the stable deformed mean field.
We show the decisive
role of the mixing between the valence spherical orbitals of the same parity
split as a function of deformation. This mixing, different for the two sides
of the axial deformation diagram, makes the prolate deformation
energetically favorable with high probability. This picture is supported by the
statistical analysis of the random interaction ensemble and by singling out the
responsible components of the effective interaction. The process
is amplified by the presence of two kinds of nucleons. Although
we have shown in detail in this Letter only cases of simple
configurations, the results are quite generic for a small system. The effect is
driven by the surface orbitals but it still remains to
see if it indeed disappears in the macroscopic limit as predicted in
Ref. \cite{hamamoto09}. The random interaction ensemble provides a new powerful tool
for understanding the many-body mechanisms of collective phenomena.

Support from the NSF grant PHY-0758099 is acknowledged. V.Z. is
grateful to S. Frauendorf, I. Hamamoto, B. Mottelson and A. Volya for useful
discussions.


\vspace*{-6mm}

\begin{thebibliography}{99}
\vspace*{-4mm}

\bibitem{BM89} A. Bohr and B.R. Mottelson, {\sl Nuclear Structure},
vol. 2 (Benjamin, New York, 1989).

\bibitem{mottprivate} B. Mottelson, {\sl private communication}.

\bibitem{lemmer60} R.H. Lemmer, Phys. Rev. {\bf 117}, 1551 (1960).

\bibitem{nilsson55} S.G. Nilsson, Kgl. Dan. Vid. Selsk. Mat.-fys. Medd.
{\bf 29}, No. 16 (1955).

\bibitem{castel76} B. Castel and K. Goeke, Phys. Rev. C {\bf 13},
1765 (1976).

\bibitem{castel90} B. Castel, D.J. Rowe, and L. Zamick, Phys. Lett.
B {\bf 236}, 121 (1990).

\bibitem{frisk90} H. Frisk, Nucl. Phys. {\bf A511}, 309 (1990).

\bibitem{deleplanque04} M.A. Deleplanque {\sl et al.},
Phys. Rev. C {\bf 69}, 044309 (2004).

\bibitem{arita98} K. Arita, A. Sugita, and K. Matsuyanagi, Czechoslovak J.
Phys. {\bf 48}, 821 (1998).

\bibitem{hamamoto91} I. Hamamoto, B.R. Mottelson, H. Xie, and X.Z. Zhang,
Zeitschr. Phys. D {\bf 21}, 163 (1991).

\bibitem{pollen03} J. Martin, M. Torrell, A.A. Korobkov, and J. Valles,
Plant Biol. No. 5, 85 (2003).

\bibitem{hamamoto09} I. Hamamoto and B.R. Mottelson, Phys. Rev. C {\bf 79},
034317 (2009).

\bibitem{tajima02} N. Tajima, Y.R. Shimizu, and N. Suzuki, Prog. Theor. Phys.
Suppl. {\bf 146}, 628 (2002).

\bibitem{brody81} T.A. Brody {\sl et al.},
Rev. Mod. Phys. {\bf 53}, 385 (1981).

\bibitem{big} V. Zelevinsky,
et al.,
B.A. Brown, N. Frazier, and M. Horoi,
Phys. Rep. {\bf 276}, 85 (1996).

\bibitem{johnson98} C.W. Johnson, G.F. Bertsch, and D.J. Dean,
Phys. Rev. Lett. {\bf 80}, 2749 (1998).

\bibitem{HBZ01} M. Horoi, B.A. Brown, and V. Zelevinsky, Phys.
Rev. Lett. {\bf 87}, 062501 (2001).

\bibitem{chau02} P. Chau Huu-Tai, A. Frank, N.A. Smirnova, and P. Van
Isacker, Phys. Rev. C {\bf 66}, 061302 (2002).

\bibitem{bijker02} R. Bijker and A. Frank, Phys. Rev. C {\bf 65},
044316 (2002).

\bibitem{mulhall00} D. Mulhall, A. Volya, and V. Zelevinsky, Phys.
Rev. Lett. {\bf 85}, 4016 (2000).

\bibitem{papenbrock05} T. Papenbrock and H.A. Weidenm\"{u}ller,
Nucl. Phys. {\bf A757}, 422 (2005).

\bibitem{ZV04} V. Zelevinsky and A. Volya, Phys. Rep. {\bf 391},
311 (2004).

\bibitem{zhao04} Y.M. Zhao, A. Arima, and N. Yoshinaga, Phys.
Rep. {\bf 400}, 1 (2004).

\bibitem{weiden07} H.A. Weidenm\"{u}ller and G.E. Mitchell, arXiv 0807.1070.

\bibitem{johnson07} C.W. Johnson, H. Nam, Phys. Rev. C {\bf 75},
047305 (2007).

\bibitem{rmpsm} E. Caurier {\sl et al.},
Rev. Mod. Phys. {\bf 77}, 427 (2005).

\bibitem{ni56} M. Horoi {\sl et al.},
Phys. Rev. C {\bf 73}, 061305(R) (2006).

\end{thebibliography}
\end{document}